\documentclass[dvipdfm]{article}
\usepackage[dvips]{graphicx} 
\usepackage[usenames,dvipsnames]{color}
\usepackage[dvipdfm]{hyperref} 

\newcommand{\bq}{\begin{eqnarray}}
\newcommand{\eq}{\end{eqnarray}}

\begin{document}

\title{Cumulant Expansion and Monthly Sum Derivative}
\author{V.M. Belyaev\\
Allianz Investment Management, Minneapolis, MN, USA}

\maketitle
\begin{abstract}
Cumulant expansion is used to derive accurate closed-form approximation for Monthly Sum Options  in case of constant volatility model.
Payoff of Monthly Sum Option is based on sum of $N$ caped (and probably floored) returns.
It is noticed, that $1/\sqrt{N}$ can be used as a small parameter in Edgeworth expansion. First two leading terms of this expansion
are calculated  here.
It is shown that the suggest closed-form approximation  is in a good agreement with numerical results for typical mode parameters.
\end{abstract}

\section{Introduction}

Monthly Sum (MS)  Options  are embedded into popular fixed index annuities.
In contrast to plain vanilla options, the closed-form formula for MS derivative is not known even in the simplest case of constant volatility and we
need to use Monte-Carlo simulations to price them.
To derive accurate approximation for MS options we can use some sort of expansion with small parameter.
In the next section we will shown that MS derivative can be considered as a Call option on underlying with caped (and probably floored) returns.
Caped distribution is very different from Gaussian one but,
according to the central theorem, in the limit of large number of returns ($N\to\infty$), the final distribution at time  of expiration will be close to
the normal one. So, in the leading $1/\sqrt{N}$ approximation, we can use standard  European Call formula  to price MS options\cite{BS},\cite{Merton}.

To calculate $1/\sqrt{N}$ corrections
 we need to choose convenient way to approximate non-normal distributions of caped underlyings.
Various types of  expansions in
series of Hermit polynomials are used in science to approximate distributions: namely Gram-Charlier, Gauss-Hermite and Edgeworth.
But Gram-Charlier series has poor convergence properties \cite{GramCharlier}.
Gauss-Hermit expansion is better but it has no expansion parameter and therefore it has no intrinsic measure of accuracy.
The most prominent is Edgeworth expansion which has an expansion parameter  $1/\sqrt{N}$ \cite{astro}. 
 This expansion is a true asymptotic series and  has good convergence properties.

Here we consider  Black-Scholes case of constant  market volatility.
It is shown, that for 
 typical Monthly Sum option and market parameters,  first two leading terms of this expansion give a good approximation for MS derivative. 
Note, that these first two terms of cumulant expansion are identical for all three types of cumulant expansion. But next corrections need
to be grouped as terms of Edgeworth series to be a true asymptotic expansion with $1/\sqrt{N}$ as a small parameter.
Higher terms of cumulant expansion
can improve accuracy of this approximation.

\section{Monthly Sum Derivative}

Monthly Sum option (MS) is defined as a derivative which payoff at the time of expiration $t_n=T$ is
\bq
(MS\; Payoff)(cap)=Max\left(\sum_{m=1}^Nr_m(cap),0\right)
\eq
where
\bq
r_m(cap)=Min\left(S(t_m)/S(t_{m-1})-1,cap\right)
\label{r}
\eq
are caped returns.

In the case  of small returns we can use logarithmic returns to approximate returns (\ref{r}) 
\bq
r_m(cap)\simeq \tilde r_m(cap)= Min\left(\ln(S(t_m)/S(t_{m-1})),\widetilde{cap}\right)
\label{lr}
\eq
where $\widetilde{cap}=\ln(1+cap)$,
and the derivative payoff is
\bq
(MS\;Payoff)(cap)\simeq (MSLN\;Payoff)(cap)=
\nonumber
\\
=Max\left[
\exp\left(
\sum_{m=1}^N\tilde r_m(cap)\right)
-1,0
\right]
=Max\left[
\frac{\tilde S(t_N)}{S(t_0)}-1,0
\right].
\eq
where
\bq
\tilde S(t_N)=S(T_0)e^{\sum_{m=1}^N\tilde r_m(cap)}
\label{caped}
\eq
It means that we can consider MS option as a call option on caped distribution.
\bq
MS\simeq MSLN=\frac1X Call^*(X=S(t_0),t_N)
\label{call}
\eq
where $Call^*(X)$ is a call payoff on the caped underlying
$\tilde S(t_N)$

\begin{figure}
\begin{center}
\includegraphics[width=\textwidth, bb=0 0 400 200]{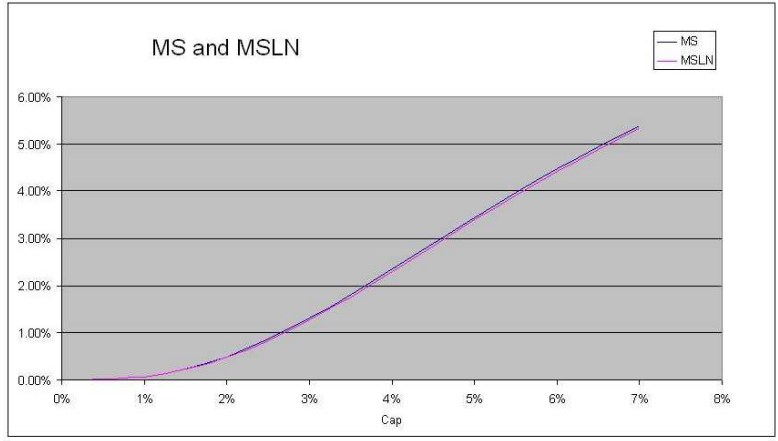}
\end{center}
  \caption{One Year MS and MSLN derivative values. Volatility=20\%, Rate=3\%, Dividend Yield=2\%, Term=1 year. 100,000 Monte-Carlo scenarios.
}
 \label{fig:ln}
\end{figure}

 Fig.\ref{fig:ln} demonstrates that, for typical market parameters, MSLN  can be used as an accurate approximation for  Monthly Sum derivative.

\section{Distribution Function and Cumulant Expansion}

To calculate call option value in eq.(\ref{call}) and therefore to calculate MS derivative, we need to know distribution of caped underlying $\tilde S(T)$ (\ref{caped}).

To approximate this distribution at expiration time, it is convenient to represent
probability distribution function of logarithmic returns at time of expiration $T$  in the form of infinite series as
\bq
\psi(x)=\frac1{\sqrt{2\pi v^2 T}}e^{-\frac{(x-\nu T)^2}{2v^2T}}\left(
1+\epsilon_1 H_3\left(\frac{x-\nu T}{v\sqrt{T}}\right)+ \dots 
\right)=
\nonumber
\\
=\psi_0\left(\frac{x-\nu T}{v\sqrt{T}}\right)+\psi_1\left(\frac{x-\nu T}{v\sqrt{T}}\right)+\dots;
\label{psi}
\eq
where $x=\ln(\tilde S(T)/S(0)$ is a logarithmic return, $\tilde S(T)$ is a price of caped underlying at time $T$,  $\psi_n(x)$ is $n$-term of  Edgeworth expansion, 
and $H_n(z)$ are Hermit polynomials:
\bq
H_n(z)=(-1)^ne^{z^2/2}\frac{d^n}{dz^n}e^{-z^2/2};
\;\;
H_3(z)=z^3-3z.
\eq
Therefore the value of Monthly Sum derivative is
\bq
MS\simeq MSLN=MS^{(0)}+MS^{(1)}+\dots
\eq
where
\bq
MS^{(n)}=e^{-rT}\int_0^\infty\left(e^x-1\right)\psi_n(x)dx
\label{int}
\eq
are terms of Edgeworth expansion for Monthly Sum option, $r$ is interest rate.

Calculating moments of distribution function $\psi$ 
we can determine parameters of 
this expansion:
\bq
\nu T= \kappa_1;\;\;\ v^2 T=\kappa_2;\;\;  6\epsilon_1\sqrt{(v^2T)^3}=\kappa_3;
\dots
\eq
where $\kappa_n$ are cumulants.

Using the fact that cumulant of sum of independent variables is equal to sum of cumulants, it is easy to calculate cumulants
of longer term distributions as
\bq
\kappa_n=N \iota_n;
\eq
where $N$ is number of time intervals (months) and $\iota_n$ is a cumulant of caped monthly distribution.

Monthly cumulants  can be expressed in terms of moments as
\bq
\iota_1=I_1;\;\; \iota_2=I_2-I_1^2;\;\; \iota_3=I_3-3I_1 I_2+2I_1^3;\;\;\dots
\label{cum}
\eq
where
\bq
I_n=\int_{-\infty}^{cap} \frac{x^ndx}{\sqrt{2\pi\sigma^2 \Delta t}}e^{-\frac{(x-\mu dt)^2}{2\sigma^2 \Delta t}}+(cap)^n\int_{cap}^\infty \frac{dx}{\sqrt{2\pi\sigma^2 \Delta t}}e^{-\frac{(x-\mu dt)^2}{2\sigma^2 \Delta t}}
\eq
are moments of caped distribution, $\Delta t=t_{n}-t_{n-1}=1/12$ is a  time interval, $\sigma$ is a market volatility,
\bq
\mu=r-y-\frac12 \sigma^2
\eq
is a risk neutral drift, $y$ is a dividend yield.

First two moments of caped distributions $I_1$ and $I_2$  give us parameters of leading order approximation for distribution function $\psi^{(0)}$ of caped underlying:
\bq
\nu T=N I_1;\;\; v^2 T=N (I_2-I_1^2).
\eq
It means, that in the leading approximation, MS derivative value is an European ATM call option 
with volatility $v$ and drift $\nu$. To set correct value for drift we need to adjust dividend yield $y$ to its effective value
\bq
\tilde y=r-\nu-\frac12 v^2
\eq
Then, the price of Monthly Sum derivative in the leading order of cumulant expansion is
\bq
MS^{(0)}=\frac1X\;Call(X+S(0),v,r,\tilde y),
\eq

\begin{figure}
\begin{center}
\includegraphics[width=\textwidth, bb=0 0 400 200]{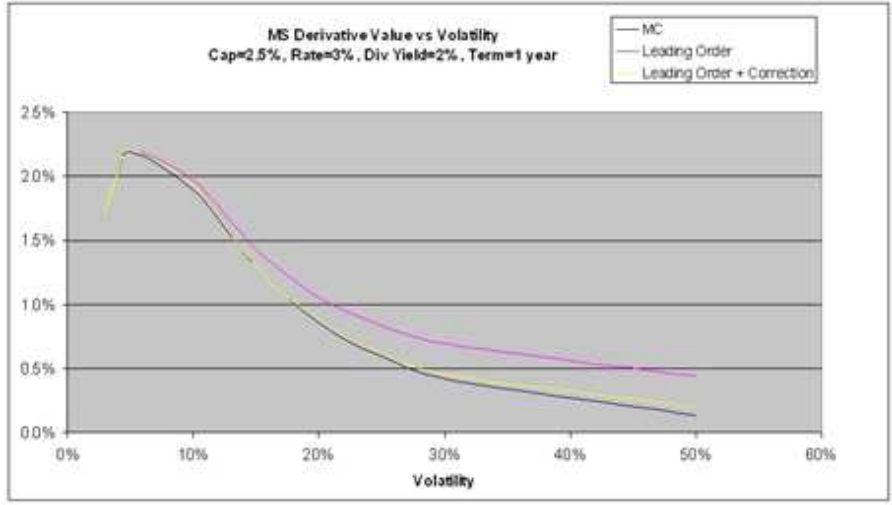}
\end{center}
  \caption{One Year MS Derivative Value. Cap=2.5\%, Rate=3\%, Div Yield=2\%, Term=1 year.
100,000  scenarios were used in Monte-Carlo calculations (MC).
}
 \label{fig:aprx}
\end{figure}

On Fig.\ref{fig:aprx} we can observe that the accuracy of leading order approximation for One Year MS option with $cap=2.5\%$  is not good   for market volatility $\sigma>10\%$.
To improve accuracy we need to  take into account higher terms of cumulant expansion (\ref{psi}).

The first correction to the leading term of cumulant expansion for probability distribution function (\ref{psi})  is
\bq
\psi_1(x)=\epsilon_1 \frac1{\sqrt{2\pi v^2 T}}e^{-\frac{(x-\nu T)^2}{2v^2T}}H_3\left(\frac{x-\nu T}{v\sqrt{T}}\right);
\eq
and according to eq.(\ref{cum}) 
\bq
\epsilon_1=N\frac{I_3-3I_1I_2+2I_1^3}{6\sqrt{v^2 T}}
\eq
As can we see on Fig.\ref{fig:aprx} this correction significantly improve accuracy of approximation.

Formulas for moments of caped distribution and integral for $M^{(n)}$ ($n=1$ eq.(\ref{int}))
 are presented in Appendix.A.

We  also can consider MS derivative for caped returns with floor:
\bq
r_m(cap,floor)=Max\left(Min\left(S(t_m)/S(t_{m-1})-1,cap\right),floor\right)
\eq
In this case, moments of this distributions can be calculated as

\bq
I_n=(floor)^n\int_{-\infty}^{floor}\frac{dx}{\sqrt{2\pi\sigma^2 \Delta t}}e^{-\frac{(x-\mu dt)^2}{2\sigma^2 \Delta t}}
+   \int_{floor}^{cap} \frac{x^ndx}{\sqrt{2\pi\sigma^2 \Delta t}}e^{-\frac{(x-\mu dt)^2}{2\sigma^2 \Delta t}}+
\nonumber
\\
+(cap)^n\int_{cap}^\infty \frac{dx}{\sqrt{2\pi\sigma^2 \Delta t}}e^{-\frac{(x-\mu dt)^2}{2\sigma^2 \Delta t}}
\label{capfloor}
\eq
First three moments of this distribution are calculated in Appendix.A eqs.(\ref{fi1}-\ref{fi3}).

\section{Conclusions}

In present paper we apply Cumulant Expansion to derive closed form approximation for Monthly Sum options.
It is shown, that for typical MS derivative and market parameters  this approximation works well. 
Calculation of higher terms of Edgeworth expansion can improve accuracy of derived approximation.

\pagebreak
\appendix
\begin{center}`
    {\bf\Large Appendices}
  \end{center}

\section{Moments of caped distributions and correction to MS price.} 
 \renewcommand{\theequation}{A.\arabic{equation}}
 \setcounter{equation}{0}
Below, we present formulas for the first three moments of caped distribution
\bq
I_n=\int_{-\infty}^\infty x^n \phi(x)dx
\eq
where
\bq
\phi(x)  =  \Theta(cap-x)\frac1{\sqrt{2\pi\sigma^2\Delta t}}e^{-\frac{(x-\mu \Delta t)^2}{2\sigma^2\Delta t}}+
 C \delta(x-cap).
\eq
 \bq
&  & C=\int_{cap}^\infty\frac{dz}{\sqrt{2\pi\sigma^2\Delta t}}e^{-\frac{(z-\mu \Delta t)^2}{2\sigma^2\Delta t}}=1-CND(\tilde c);
\nonumber
\\
&  &  \tilde c=\frac{cap-\mu\Delta t}{\sqrt{\sigma^2 \Delta t}};
\nonumber
\\
&  & CND(z)=\int_{-\infty}^z\frac{dx}{\sqrt{2\pi}}e^{-\frac12 x^2}=\frac12 \left(1+erf(x/\sqrt{2}) \right);
\nonumber
\\
&  & \Theta(z)=\left\{
\begin{array}{ll}
0 &\textrm{if $x\leq 0$}\\
1 &\textrm{if $x> 0$}
\end{array}
\right.
\eq
and $\delta(x)$ is  Dirac $\delta$-function.
\bq
I_1 & = & -\sigma e^{-\frac12\tilde c^2}\sqrt{\frac{\Delta t}{2\pi}}+\mu\; \Delta t \;CND(\tilde c)+C\; cap
\\
I_2 & = & \sigma^2 \Delta t\;CND(\tilde c)-\frac{\tilde c}{\sqrt{2\pi}}\sigma^2 \Delta t e^{-\tilde c^2}-2\mu\;\sigma\;\Delta t\;e^{-\tilde c^2/2}\sqrt{\frac{\Delta t}{2\pi}}+
\nonumber
\\
 & + &(\mu \Delta t)^2CND(\tilde c)+C\; (cap)^2
\\
I_3 & = &-\left(1+\frac12\tilde c^2\right)e^{-\tilde c^2/2}\sqrt{\frac{2(\sigma^2\Delta t)^3}{\pi}}+
\nonumber
\\
 & + &3\mu\;\Delta t\left(
\sigma^2\Delta t\; CND(\tilde c)-(cap-\mu\;\Delta t)e^{-\tilde c^2/2}\sqrt{\frac{\sigma^2\Delta t}{2\pi}}
\right)-
\nonumber
\\
& - &  3(\mu\;\Delta t)^2\sigma\;e^{-\tilde c^2/2}\sqrt{\frac{\Delta t}{2\pi}}+(\mu\;\Delta t)^3CND(\tilde c)+C\;(cap)^3
\eq

First correction to the leading approximation  (\ref{int})
has integral which can be presented in the following form:
\bq
& & \int_0^\infty\left(e^x-1\right)
\frac1{\sqrt{2\pi v^2 T}} e^{-\frac{(x-\nu T)^2}{2v^2T}}  H_3\left(
\frac{x-\nu T}{v\sqrt{T}}
\right)dx=
\nonumber
\\
& & =\frac{T}{\sqrt{2\pi}}(v^2-\nu)e^{-\frac{\nu^2 T}{v^2}}+(v^2 T)^{3/2}e^{(\nu+\frac12 v^2)T}CND\left( \left(\frac{\nu}{v}+v\right)\sqrt{T}\right)
\eq

To calculate MS derivative value with local floor, we need to calculate integrals in eq.(\ref{capfloor}) .
This integrals can be presented in the following form
\bq
I_n=(floor)^n CND(\tilde f)+(cap)^n(1-CND(\tilde c))+
\nonumber
\\
+\int_{floor}^{cap}\frac{x^ndx}{\sqrt{2\pi\sigma^2 \Delta t}}e^{-\frac{(x-\mu\Delta t)^2}{2\sigma^2\Delta t}}
\eq
where
\bq
\tilde f=\frac{floor-\mu \Delta t}{\sqrt{2\pi \sigma^2 \Delta t}}
\eq

Using that
\bq
\int_{floor}^{cap}\frac{x^ndx}{\sqrt{2\pi\sigma^2 \Delta t}}e^{-\frac{(x-\mu\Delta t)^2}{2\sigma^2\Delta t}}=
\frac{(\sigma^2\Delta t)^{n/2}}{\sqrt{2\pi}}\int_{\tilde f}^{\tilde c}(x+\tilde\mu)^n e^{-x^2/2}dx
\eq
where
\bq
\tilde\mu=\frac{\mu}{\sigma}\sqrt{\Delta t}
\eq
we obtain the following formulas for the first three moments of the caped-floored distribution
\bq
I_1 & = & \sigma\left(
e^{-\tilde f^2/2}-e^{-\tilde c^2/2}
\right)\sqrt{\frac{\Delta t}{2\pi}}+\mu\Delta t( CND(\tilde c)-CND(\tilde f))+
\nonumber
\\
 & + & floor\; CND(\tilde f)+cap\;(1- CND(\tilde c))
\label{fi1}
\\
I_2 & = & \sigma^2\Delta t(CND(\tilde c)-CND(\tilde f))
-\frac{\sigma^2\Delta t}{\sqrt{2\pi}}\left(
\tilde c e^{-\tilde c^2/2}-\tilde f e^{-\tilde f^2/2}
\right)+
\nonumber
\\
& + & 2\mu\sigma\Delta t\left(
e^{-\tilde f^2}-e^{-\tilde c^2}
\right)\sqrt{\frac{\Delta t}{2\pi}}+
\nonumber
\\
& + &(\mu\Delta t)^2(CND(\tilde c)-CND(\tilde f))+
\nonumber
\\
& + & (cap)^2(1-CND(\tilde c))+(floor)^2CND(\tilde f)
\label{fi2}
\\
I_3 & = & -\left(
\left(2+\tilde c^2\right)e^{-\tilde c^2/2}-\left(2+\tilde f^2\right)e^{-\tilde f^2/2}
\right)\sqrt{\frac{(\sigma^2 \Delta t)^3}{2\pi}}+
\nonumber
\\
& + & 3\mu(\sigma\Delta t)^2\left(
CND(\tilde c)-CND(\tilde f)-\frac1{\sqrt{2\pi}}\left(\tilde c e^{-\tilde c^2/2}-\tilde f  e^{-\tilde f^2/2}
\right)
\right)-
\nonumber
\\
& - & 3 (\mu\Delta t)^2\sigma\left(
e^{-\tilde c^2/2}-e^{-\tilde f^2/2}
\right)\sqrt{\frac{\Delta t}{2\pi}}+
\nonumber
\\
& + &  (\mu\Delta t)^3(CND(\tilde c)-CND(\tilde f)) +
\nonumber
\\
& + & (cap)^3(1-CND(\tilde c))+(floor)^3CND(\tilde f)
\label{fi3}
\eq
\end{document}